\documentclass[11pt]{article}
   
\usepackage{epsfig}
\usepackage{amsfonts,amssymb}

\renewcommand{\theequation}{\arabic{section}.\arabic{equation}}

\newcommand{\del}{\ensuremath{\partial}}

\newcommand{\half}{\ensuremath{\frac{1}{2}}}

\newcommand{\be}{\begin{equation}}
\newcommand{\ee}{\end{equation}}

\newcommand{\ba}{\begin{eqnarray}}
\newcommand{\ea}{\end{eqnarray}}

\newcommand{\ns}{\normalsize}

\begin{document}


\begin{titlepage}

\title{
   \hfill{\ns hep-th/0506138\\}
   \vskip 2cm
   {\Large\bf A practical model for cosmic (p,q) superstrings.}
\\[0.5cm]}
   \setcounter{footnote}{0}
\author{
{\ns\large 
  \setcounter{footnote}{3}
  P. M. Saffin\footnote{email: P.M.Saffin@sussex.ac.uk}}
\\[0.5cm]
   {\it\ns Department of Physics and Astronomy, University of Sussex}
   \\
   {\ns Falmer, Brighton BN1 9QJ, UK} \\[0.2em] }
\date{}

\maketitle

\begin{abstract}\noindent
We propose a model with a U(1)$_A\times$U(1)$_B$ gauge symmetry which contains topological
strings carrying magnetic flux under each of the U(1)s. 
By calculating the tension of the first few low energy strings we show that bound
states, containing flux under each of the U(1)s, are stable against decay to their 
constituents of lower winding number. As the model contains only Abelian gauge symmetries this model
provides a pragmatic solution to the numerical modelling of cosmic-superstrings.
Whilst these defects do not satisfy a BPS bound we argue that they bear sufficient similarities
to warrant such a study.
\end{abstract}

\thispagestyle{empty}

\end{titlepage}

\section{Introduction}
\label{sec:introduction}
Cosmic strings have come in and out of fashion a number times, and whilst it
is clear that they do not play a lead role in structure formation
they may nevertheless have observational consequences
\cite{Contaldi:1998qs,Jeannerot:2003qv,Wyman:2005tu}. Indeed, it is
hoped that advanced LIGO and LISA will be able to detect the characteristic radiation
of a string network
\cite{Damour:2000wa,Damour:2001bk,Damour:2004kw}.
The context of the current revival of cosmic strings dates back to work of Witten
in perturbative string theory \cite{Witten:1985fp} where it was concluded that
superstrings could not act as cosmic strings. Since then it has become clear that
there are more possibilities for the compact dimensions of string theory, which has
led to the issue of cosmic superstrings being reconsidered
\cite{Jones:2002cv,Sarangi:2002yt,Copeland:2003bj}; for a nice overview
see \cite{Polchinski:2004ia}. The particular networks that we are interested in are
associated to the low energy dynamics of type IIB string theory which, in ten dimensions,
contains two different species of string. First of all there is the fundamental, F-string
which carries charge under the Neveu Schwartz-Neveu Schwartz two-form potential, 
secondly there is the
Dirichlet, D-string carrying charge under the Ramond-Ramond two-form potential.
The types of string that motivate the current work are $(p,q)$ strings which are
a supersymmetric bound state of F and D strings.
These come about because parallel F and D strings are not supersymmetric
and can reduce their energy by the F string breaking on the D string with the endpoints
moving off to infinity \cite{Li:1995pq,Polchinski:1996na}, the resulting
configuration is then supersymmetric and has the following tension,
\ba
\label{eqn:pqTension}
\mu_{p,q}=\sqrt{(\mu_p)^2+(\mu_q)^2}=\mu_F\sqrt{p^2+q^2/g_s^2}.
\ea
where $\mu$ refers to the tension and $g_s$ is the string coupling.
When we consider the physics of four dimensions we note that there are more general ways
to get string solutions by utilizing the branes of string theory. If we were to wrap
n-1 of the spatial dimensions of an n-brane around a compact cycle then such an object
would appear as a string in four dimensions, with couplings and tensions that depended
upon the details of the compactification. One model which has been worked on in detail
is the \mbox{\it K\hspace{-7pt}KLM\hspace{-9pt}MT} model \cite{Kachru:2003sx}, using
a warped compactification with matter fields living on a D3 brane in the throat. The tension
of objects is then affected by the warping, with the redshift factor allowing for a
large range of effective string tension. In principle there can be many such throats,
each with a different warping, this would then produce strings with a variety of
different tensions. Modelling such situations may well follow similar lines to those
put forward here, adding an extra U(1) for each species.

Modelling the evolution of a $(p,q)$ network involves different approaches to that of the
standard Nielsen-Olesen string due to the way $(p,q)$ re-connect as two strings pass through
each other. Abelian strings simply swap partners as they interact
\cite{Shellard:1987bv,Copeland:1986ng,matzner:1988,Hanany:2005bc}
but this channel is not open to $(p,q)$ strings as this would violate charge conservation.
For such partner-swapping to take place an intermediate
string is created, joining together the two initial strings in much the same way as for
non-Abelian gauge vortices \cite{mermin:1979,Hashimoto:2005hi}.
This raises the possibility that these networks get more tangled as they evolve, eventually
freezing out as the Universe continues to expand. This would cause the string network
to eventually dominate the energy density of the Universe \cite{Spergel:1996ai,Bucher:1998mh}.
An important issue therefore is to decide whether these networks reach a scaling solution
\cite{Kibble:1976sj}, meaning that the string energy density forms a constant
fraction of the background energy density, with the network length scale increasing
proportional to time.
Analytic and numerical approaches are now being developed to approach this problem
\cite{Tye:2005fn,Copeland:2005cy,Martins:2004vs}, here we present a simple model 
containing gauged vortices which has many of the required properties, moreover it
is well suited to numerical implementation by adapting the available codes for evolving
Abelian gauge vortices. This is in counter-distinction to modelling these networks with
non-Abelian gauge field theories, where imposing lattice gauge symmetry is numerically
intensive \cite{McGraw:1997nx}.

The issue of the correct field theory to use when describing the low energy physics
of D-strings in four dimensions is now converging to a supersymmetric model containing:
chiral fields transforming in the usual way under U(1) gauge transformations;
a chiral field, representing an axion, with a shift symmetry that gets gauged under 
the U(1); and a vector multiplet to carry the magnetic flux of the vortex\cite{Blanco-Pillado:2005xx}.
This builds on earlier work where the axion multiplet was not included
\cite{Dvali:2003zh,Binetruy:1998mn,Achucarro:2004ry}. An obvious drawback of the
model put forward here is that it is not based on a supersymmetric field theory. However, the
simple form of coupling between the two U(1) sectors 
in our model leads us to believe that it
should be possible to connect two supersymmetric U(1) gauge theories to yield a theory
with vortex bound states of the type described below, whether such vortices would be BPS remains
to be seen.

We organize the paper by first introducing the model in section \ref{sec:model} and describing 
how the vacuum manifold depends on the parameters. The parameter space is then reduced by
considering the BPS limit of the theory in section \ref{sec:BPS}. In section \ref{sec:solutions}
we construct the ansatz for the vortices and calculate their energy, we then draw our
conclusions in section \ref{sec:conclusions}.

\section{the model}
\label{sec:model}
The model is based on the simplest way to couple two independent U(1) gauge theories and
has been seen before, albeit in a different guise \cite{Witten:1984eb,Davis:1988jq}. 
There it was used
to construct cosmic strings which could form a condensate in their core and become superconducting.
For that to work the parameters are chosen such that one of the Higgs fields acquires a
non-zero vacuum expectation value (vev) while the other sits at zero; here we require
the ``opposite''. What we need is for the vacuum to be such that {\it both} Higgs fields acquire
a vev, allowing for there to be independent string solutions following the usual
arguments \cite{vilenkin:1994}. 
Although our action is the same as those for the superconducting string models, because
we use the opposite parameter regime we write it in a way more suited to our analysis
\ba
\label{eqn:lagrangian}
{\cal L}&=&-D_\mu\bar\phi D^\mu\phi-{\cal D}_\mu\bar\psi{\cal D}^\mu\psi
-\frac{1}{4}F_{\mu\nu}F^{\mu\nu}
-\frac{1}{4}{\cal F}_{\mu\nu}{\cal F}^{\mu\nu}
-V(|\phi|,|\psi|).
\ea
Where we have the following gauge-covariant derivatives, field strengths and potential,
\ba
D_\mu&=&\del_\mu-ieA_\mu,\qquad{\cal D}_\mu=\del_\mu-igB_\mu,\\
F_{\mu\nu}&=&\del_\mu A_\nu-\del_\nu A_\mu,\qquad {\cal F}_{\mu\nu}=\del_\mu B_\nu-\del_\nu B_\mu,\\
\label{eqn:potential1}
V(|\phi|,|\psi|)&=& \frac{\lambda_1}{4}\left(\bar\phi\phi-\eta^2\right)^2
                   +\frac{\lambda_2}{4}\left(\bar\psi\psi-\nu^2\right)^2
                   -\kappa\left(\bar\phi\phi-\eta^2\right)\left(\bar\psi\psi-\nu^2\right).
\ea
For generic values of the coupling constants we find that the critical ``points'' are given by,
\ba
(\bar\phi\phi,\bar\psi\psi)_0&=&(0,0),\quad (\eta^2,\nu^2),\quad (\eta^2-\frac{2\kappa}{\lambda_1}\nu^2,0)
                                ,\quad(0,\nu^2-\frac{2\kappa}{\lambda_2}\eta^2).
\ea
When constructing the superconducting strings it was the last two critical points that were
of interest, as these have one broken and one unbroken U(1) symmetry, and so the parameters
were chosen to make these minima.
As we are going to need two broken U(1)s then we need to make the critical point $(\eta^2,\nu^2)$
a minimum.
To gain further insight on
the vacuum structure we complete the square as follows
\ba
V(|\phi|,|\psi|)&=& \frac{1}{4}\left[\sqrt{\lambda_1}(\bar\phi\phi-\eta^2)\pm\sqrt{\lambda_2}(\bar\psi\psi-\nu^2)\right]^2
                   -[\kappa\pm\half\sqrt{\lambda_1\lambda_2}](\bar\phi\phi-\eta^2)(\bar\psi\psi-\nu^2),
\ea
and note that there are special values of $\kappa$ where the manifold of minima increases
in dimension. For $\kappa=-\half\sqrt{\lambda_1\lambda_2}$ we get a vacuum which is an ellipsoid,
\ba
\sqrt{\lambda_1}\bar\phi\phi+\sqrt{\lambda_2}\bar\psi\psi=\sqrt{\lambda_1}\eta^2+\sqrt{\lambda_2}\nu^2,
\ea
while for $\kappa=+\half\sqrt{\lambda_1\lambda_2}$ we get a vacuum which is a hyperboloid
\ba
\sqrt{\lambda_1}\bar\phi\phi-\sqrt{\lambda_2}\bar\psi\psi=\sqrt{\lambda_1}\eta^2-\sqrt{\lambda_2}\nu^2.
\ea
It is these special values of $\kappa$ which separate the different classes of vacuum structure.
For $\kappa<-\half\sqrt{\lambda_1\lambda_2}$, the minima are at $\quad (\eta^2-\frac{2\kappa}{\lambda_1}\nu^2,0)$,
$(0,\nu^2-\frac{2\kappa}{\lambda_2}\eta^2)$, which is the parameter range for superconducting strings.
For $-\half\sqrt{\lambda_1\lambda_2}<\kappa<+\half\sqrt{\lambda_1\lambda_2}$, the minima are at 
$(\eta^2,\nu^2)$, which is the range we are interested in.
For $\kappa>+\half\sqrt{\lambda_1\lambda_2}$ the potential is unbounded below and so is unphysical.
Within our range of interest, $-\half\sqrt{\lambda_1\lambda_2}<\kappa<+\half\sqrt{\lambda_1\lambda_2}$ there is another
special value of $\kappa$, namely zero. For $\kappa=0$ we see that the two U(1) are entirely decoupled
and so the vortices of U(1)$_A$ do not talk to those of U(1)$_B$, in that sense we could think of
two different vortex species as being neutrally bound. It is therefore natural to expect that $\kappa=0$
is a boundary between the two vortex species having negative or positive binding energy, i.e. forming
bound states or not. To see which sign of $\kappa$ gives bound states consider two separated vortices,
one of each species. At the $A$ vortex we have $\phi=0$, $|\psi|\lessapprox\nu$ while at the $B$ vortex we $\psi=0$,
$|\phi|\lessapprox\eta$. Now consider the $A$ vortex, we note from (\ref{eqn:potential1}) that for $\kappa>0$
we can lower the potential by reducing $|\psi|$, this can be achieved by bringing the vortices closer,
i.e. they attract each other.
Putting together all of the above arguments leads us to the following parameter regime for bound $A-B$
vortices
\ba
0<\kappa<\half\sqrt{\lambda_1\lambda_2}.
\ea
\subsection{BPS considerations}
\label{sec:BPS}
Although we understand the parameter space there are still a large number of parameters, and if
we are to make any progress we need to make some choices. To make further restrictions we
take our inspiration from the Bogomol'nyi limit of vortices, \cite{Bogomolny:1975de,vilenkin:1994}. 
Adapting those arguments to our case with
$\kappa=0$ the Hamiltonian takes the particular
form of a sum of squares and boundary terms if we choose $\lambda_1=2e^2$ and $\lambda_2=2g^2$.
With that, one then sees that for a state with winding number $n$ in U(1)$_A$ and $m$ in
U(1)$_B$ the minimum energy configuration has
\ba
\label{eqn:BPSvortex}
E(\kappa=0,n,m)&=&2\pi\eta^2n+2\pi\nu^2m.
\ea
Where we have made the further parameter choice $e=g=1$. 
In order to make a comparison between the bound-state energy of our system and (\ref{eqn:pqTension})
we introduce the following quantity,
\ba
\label{eqn:boundState}
\epsilon_{(n,m)}=\sqrt{(2\pi\eta^2)^2+(2\pi\nu^2)^2}.
\ea

What we shall now do is explore the remaining parameter space ($\eta$, $\nu$ and $\kappa$) to
see how the mass of $(n,m)$ vortices compares to $(p,q)$ strings. The reason we keep $\eta$ and
$\nu$ is that, from (\ref{eqn:BPSvortex}), they are expected to govern the tension of the basic constituents with
$\mu_F\sim 2\pi\eta^2$ and $\mu_F/g_s\sim 2\pi\nu^2$.

\section{Vortex solutions}
\label{sec:solutions}
What we shall now do is proceed to an evaluation of the tension of some low lying bound states.
To be specific we shall look at two cases: firstly we consider $\eta=\nu$; secondly we look 
at $\nu=2\eta$ in order to give a mass separation between the $A$ and $B$ vortices, just
as fundamental and Dirichlet strings have different tensions.

The ansatz we use for a static vortex with winding numbers $(n,m)$ is as follows,
\ba
\phi&=&\eta f(r) \exp(in\theta),\\
A_\theta&=&\frac{n}{e}\alpha(r),\\
\psi&=&\nu p(r) \exp(im\theta),\\
B_\theta&=&\frac{m}{e}\beta(r),
\ea
for which the equations of motion following from (\ref{eqn:lagrangian}) are
\ba
\label{eqn:f}
f''+\frac{1}{r}f'-\frac{n^2}{r^2}f(\alpha-1)^2-\half\lambda_1\eta^2(f^2-1)f+\kappa\nu^2(p^2-1)f&=&0,\\
p''+\frac{1}{r}p'-\frac{m^2}{r^2}p(\beta-1)^2-\half\lambda_2\nu^2(p^2-1)p+\kappa\eta^2(f^2-1)p&=&0,\\
\alpha''-\frac{1}{r}\alpha'-2e^2f^2\eta^2(\alpha-1)&=&0,\\
\label{eqn:beta}
\beta''-\frac{1}{r}\beta'-2g^2p^2\nu^2(\beta-1)&=&0.
\ea
Note that we have kept $e$, $g$, $\lambda_1$, $\lambda_2$ in these equations even though we shall 
only be using the values as described above, $e=g=\half\lambda_1=\half\lambda_2=1$.
The asymptotic boundary conditions are that $f(r\rightarrow\infty)\rightarrow 1$,
$p(r\rightarrow\infty)\rightarrow 1$, $\alpha(r\rightarrow\infty)\rightarrow 1$,
$\beta(r\rightarrow\infty)\rightarrow 1$, necessary for a finite energy configuration.
The boundary conditions at the vortex core depend on the winding number;
if $\phi$ has a non-zero winding then $f(0)=0$ and $\alpha(0)=0$ otherwise 
$f'(0)=0$ and $\alpha(0)=1$. Similar conditions hold for $\psi$.

\subsection{Case I, $\eta=\nu$}
\begin{figure}
\center
\epsfig{file=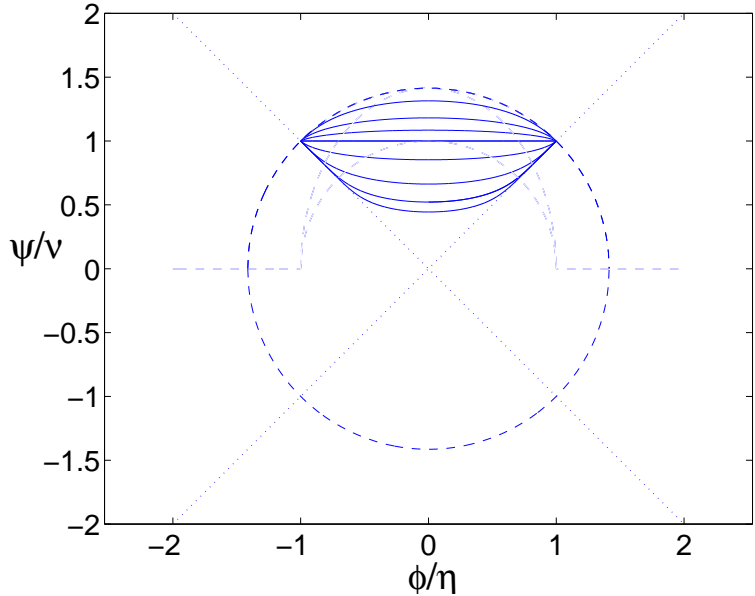,width=10cm}
\flushleft
\caption{
A representation of the field space, with the vortex profile functions
for a $(1,0)$ vortex marked. The circle is the $\kappa=-1$ vacuum and the
45$^\circ$ lines are the $\kappa=1$ vacuum.
}
\label{fig:vacuum}
\end{figure}

The first example is for $\eta=\nu$ and we can use our understanding of the potential
gained earlier to see how the profile functions should behave. In Fig. \ref{fig:vacuum}
we have a representation of the vacuum with: the points $(\pm 1,\pm 1)$ and $(\pm 1, \mp 1)$
being the vacua for $\kappa$
in our region of interest, $-1<\kappa<1$; the circle is the vacuum for $\kappa=-1$; and
the 45$^\circ$ lines are the vacuum for $\kappa=1$. On the diagram we have shown how
the Higgs field profile functions vary as we move from one vacuum at $x\rightarrow-\infty$
to another at $x\rightarrow\infty$
for a vortex with winding in the $\phi$ field but not $\psi$.
The various solid lines correspond to the vortex solutions for different values of $\kappa$,
the straight line at $\psi/\nu=1$ is the $\kappa=0$ case where $\phi$ and $\psi$ are
decoupled. The lines below this case are for $\kappa$ increasing, and they approach the
45$^\circ$ lines as $\kappa$ nears unity. If we decrease $\kappa$ below zero then the Higgs
profile functions mark out a curve in $\phi-\psi$ space which
tends toward the circle, which is the vacuum for $\kappa=-1$.

\begin{figure}
\center
\epsfig{file=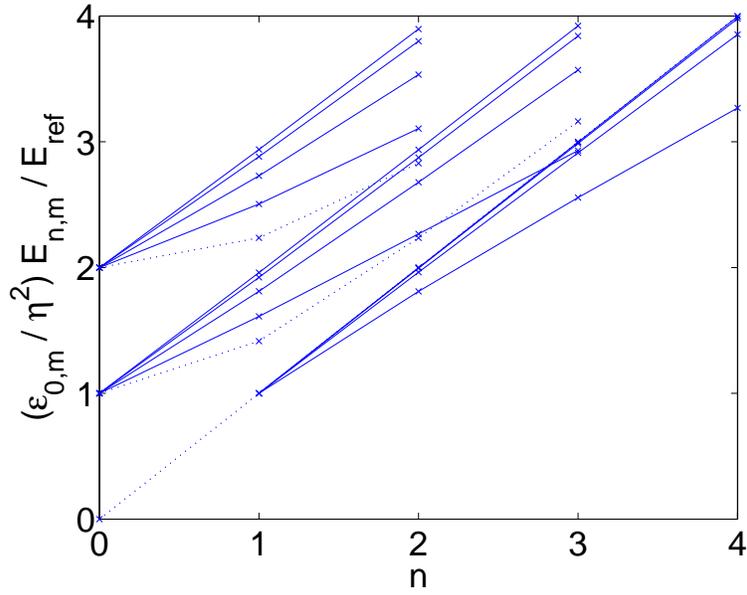,width=10cm}
\flushleft
\caption{
string tension for the case $\eta^2=\nu^2$, $\kappa$=0.1, 0.2, 0.5, 0.9. Of the three
``rays'', the lower corresponds to $m=0$, the middle has $m=1$ and the upper
has $m=2$. $E_{ref}$ is a reference energy, $E_{ref}(m=0)=E_{(1,0)}$, $E_{ref}(m>0)=E_{(0,1)}$.
}
\label{fig:tension11}
\end{figure}

In order to solve (\ref{eqn:f}-\ref{eqn:beta}) we used a numerical approach described in
\cite{numrec:1992} called successive over-relaxation. We tested our numerics against the analytic
BPS solution and we are confident that the energies quoted in the appendices are correct to
four decimal places. What is clear from table 1
is that vortices with higher
winding numbers do indeed form bound states, with the bound state energies being lower than
the sum of their constituents. Moreover, as $\kappa$ is increased toward unity (the upper
bound on $\kappa$) the binding of vortices is stronger. The results of table 1
have been plotted in Fig. \ref{fig:tension11} with the energies scaled in order to make 
comparison with (\ref{eqn:boundState}), which has been shown with dotted lines. The plot
consists of three ``rays'' corresponding to different values of $m$, with the lowest such
ray being $m=0$ and the upper being $m=2$. Within each ray we show how the tension depends on
$\kappa$, with the upper line in each ray being $\kappa=0.1$ and the lowest being $\kappa=0.9$.
We can see from this plot that the bound state spectrum does not follow the $(p,q)$ tension
spectrum of (\ref{eqn:boundState}) particularly well, but for $\kappa$ near its upper bound
it is closest.  In a real network we would expect there to be fewer high winding strings as they
are heavier, as such, taking $\kappa$ close to unity may well give a suitable approximation.

\subsection{Case II, $2\eta=\nu$}
\begin{figure}
\center
\epsfig{file=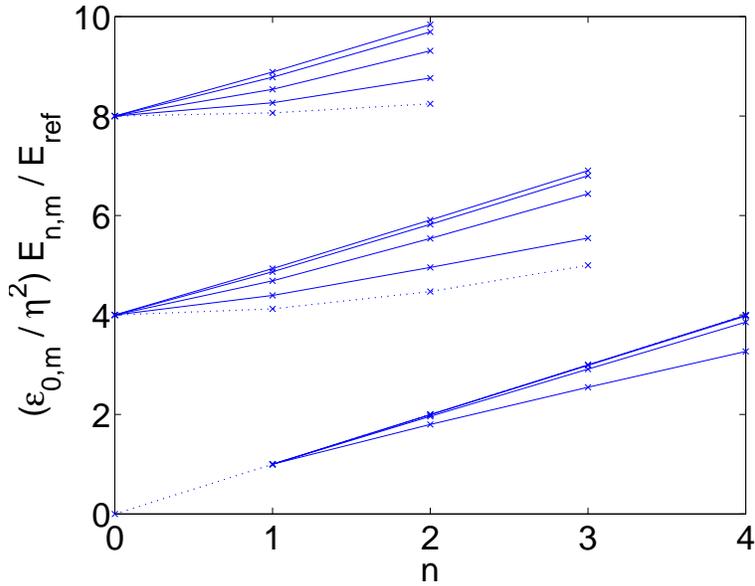,width=10cm}
\flushleft
\caption{
string tension for the case $4\eta^2=\nu^2$, $\kappa$=0.1, 0.2, 0.5, 0.9. Of the three
``rays'', the lower corresponds to $m=0$, the middle has $m=1$ and the upper
has $m=2$. $E_{ref}$ is a reference energy, $E_{ref}(m=0)=E_{(1,0)}$, $E_{ref}(m>0)=E_{(0,1)}$.
}
\label{fig:tension14}
\end{figure}

In order to represent the difference in tension between a D-string and and F-string we can 
choose to give different values to $\eta$ and $\nu$. For the purposes of illustration we took
$\nu^2=4\eta^2$ and repeated the above process to produce table 2,
which contains
the tension of the vortices for the first few windings. Again we see that higher winding
strings are bound states, with an energy lower than the sum of their constituent parts.
Fig. \ref{fig:tension14} illustrates the data in table 2, 
with the dotted lines representing (\ref{eqn:boundState}). We note that the specific tension
formula for $(p,q)$ strings seems to be modelled better in this second case, where the basic
strings have different tensions, for $\kappa$ close to its upper bound. This means that simulations
using different values for $\eta$ and $\nu$ will more closely model the dynamics of a $(p,q)$
network. 

\section{Conclusion}
\label{sec:conclusions}
In trying to model the dynamics of a network of $(p,q)$ strings coming from
string theory one must try both analytic and numerical approaches
\cite{Tye:2005fn,Copeland:2005cy,Martins:2004vs}, each with its
advantages and disadvantages. The properties of inter-commutation of $(p,q)$ strings,
i.e. the presence of three-string vertices, makes it natural model these
networks with a non-Abelian gauge theory containing a rich spectrum of vortices.
Unfortunately such simulations are extremely demanding in terms of cpu hours,
making large scale simulations intractable. As one needs a large dynamic range
in order to observe scaling behaviour this presents us with a problem; what we
have presented here is a practical resolution to this impasse. By using two
Abelian gauge symmetries it is a simple matter to adapt existing cosmic string
code to the study of $(p,q)$ networks.

An important matter that this work does not reveal is the details of the
interaction between two vortices of different type. In particular we have not
shown that such vortices inter-commute to create string junctions; it is possible
that they simply pass through each other as they would for $\kappa=0$. However,
as string junctions certainly exist in this model we expect inter-commutation
to occur. 

A natural follow-up of this work would be to try to extend the proposed
model incorporating the ideas of \cite{Blanco-Pillado:2005xx}. In \cite{Blanco-Pillado:2005xx}
it was argued that D-strings can be modelled by a U(1) supersymmetric gauge field theory
containing a chiral axion field as well as the usual chiral Higgs fields. 
If we are to have a field theory description of $(p,q)$ strings along the lines
presented here one would need to couple together two such field theories.
It is certainly possible to couple two such field theories together, however
it is unclear that one would find supersymmetric bound-states of vortices
as a solution.

\vspace{1cm}
\noindent
{\large\bf Acknowledgements} The author would like to thank 
Mark Hindmarsh for various suggestions and PPARC for financial support.


\vskip 1cm
\appendix{\noindent\Large \bf Appendices}
\renewcommand{\theequation}{\Alph{section}.\arabic{equation}}
\setcounter{equation}{0}

\section{Data for the cases $\nu^2=\eta^2$, $\nu^2=4\eta^2$.}
\label{AppA}

\begin{table}[ht!]
\label{tbl:table1}
\center
 \begin{tabular}{||l|c|c|c|c||} \hline
   $\kappa$ & $n$ & $m$ & $E_{(n,m)}$ & $nE_{(1,0)}+mE_{(0,1)}$     \\ \hline
   0.1      & 1   & 0   & 0.9985      & 0.9985     \\ \hline
   0.1      & 2   & 0   & 1.9956      & 1.9952     \\ \hline
   0.1      & 3   & 0   & 2.9922      & 2.9928     \\ \hline
   0.1      & 4   & 0   & 3.9886      & 3.9904     \\ \hline
   0.1      & 1   & 1   & 1.9568      & 1.9952     \\ \hline
   0.1      & 2   & 1   & 2.9321      & 2.9928     \\ \hline
   0.1      & 3   & 1   & 3.9160      & 3.9904     \\ \hline
   0.1      & 2   & 2   & 3.8875      & 3.9904     \\ \hline
   0.2      & 1   & 0   & 0.9939      & 0.9939     \\ \hline
   0.2      & 2   & 0   & 1.9824      & 1.9878     \\ \hline
   0.2      & 3   & 0   & 2.9688      & 2.9817     \\ \hline
   0.2      & 4   & 0   & 3.9540      & 3.9756     \\ \hline
   0.2      & 1   & 1   & 1.9098      & 1.9878     \\ \hline
   0.2      & 2   & 1   & 2.8565      & 2.9817     \\ \hline
   0.2      & 3   & 1   & 3.8184      & 3.9756     \\ \hline
   0.2      & 2   & 2   & 3.7661      & 3.9756     \\ \hline
   0.5      & 1   & 0   & 0.9586      & 0.9586     \\ \hline
   0.5      & 2   & 0   & 1.8816      & 1.9172     \\ \hline
   0.5      & 3   & 0   & 2.7910      & 2.8758     \\ \hline
   0.5      & 4   & 0   & 3.6931      & 3.8344     \\ \hline
   0.5      & 1   & 1   & 1.7358      & 1.9172     \\ \hline
   0.5      & 2   & 1   & 2.5677      & 2.8758     \\ \hline
   0.5      & 3   & 1   & 3.4233      & 3.8344     \\ \hline
   0.5      & 2   & 2   & 3.3243      & 3.8344     \\ \hline
   0.9      & 1   & 0   & 0.7931      & 0.7931     \\ \hline
   0.9      & 2   & 0   & 1.4354      & 1.5862     \\ \hline
   0.9      & 3   & 0   & 2.0274      & 2.3793     \\ \hline
   0.9      & 4   & 0   & 2.5928      & 3.1724     \\ \hline
   0.9      & 1   & 1   & 1.2780      & 1.5862     \\ \hline
   0.9      & 2   & 1   & 1.7981      & 2.3793     \\ \hline
   0.9      & 3   & 1   & 2.3235      & 3.1724     \\ \hline
   0.9      & 2   & 2   & 2.2284      & 3.1724     \\ \hline
 \end{tabular}
\caption{Table showing the energy of the vortices for $\nu=\eta$.}
\label{tab:energies1}
\end{table}

\begin{table}
\label{tbl:table2}
\center
 \begin{tabular}{||l|c|c|c|c||} \hline
   $\kappa$ & $n$ & $m$ & $E_{(n,m)}$ & $nE_{(1,0)}+mE_{(0,1)}$     \\ \hline
   0.1      & 1   & 0   & 0.9981      & 0.9981     \\ \hline
   0.1      & 2   & 0   & 1.9949      & 1.9962     \\ \hline
   0.1      & 3   & 0   & 2.9912      & 2.9943     \\ \hline
   0.1      & 4   & 0   & 3.9874      & 3.9924     \\ \hline
   0.1      & 0   & 1   & 3.9960      & 3.9960     \\ \hline
   0.1      & 1   & 1   & 4.9275      & 4.9941     \\ \hline
   0.1      & 2   & 1   & 5.9049      & 5.9922     \\ \hline
   0.1      & 3   & 1   & 6.8949      & 6.9903     \\ \hline
   0.1      & 0   & 2   & 7.9887      & 7.992     \\ \hline
   0.1      & 1   & 2   & 8.8697      & 8.9901     \\ \hline
   0.1      & 2   & 2   & 9.8226      & 9.9882     \\ \hline
   0.2      & 1   & 0   & 0.9924      & 0.9924     \\ \hline
   0.2      & 2   & 0   & 1.9794      & 1.9848     \\ \hline
   0.2      & 3   & 0   & 2.9646      & 2.9772     \\ \hline
   0.2      & 4   & 0   & 3.9490      & 3.9696     \\ \hline
   0.2      & 0   & 1   & 3.9845      & 3.9845     \\ \hline
   0.2      & 1   & 1   & 4.8498      & 4.9769     \\ \hline
   0.2      & 2   & 1   & 5.7993      & 5.9693     \\ \hline
   0.2      & 3   & 1   & 6.7722      & 6.9617     \\ \hline
   0.2      & 0   & 2   & 7.9524      & 7.969     \\ \hline
   0.2      & 1   & 2   & 8.7291      & 8.9614     \\ \hline
   0.2      & 2   & 2   & 9.6335      & 9.9538     \\ \hline
   0.5      & 1   & 0   & 0.9480      & 0.9480     \\ \hline
   0.5      & 2   & 0   & 1.8598      & 1.896     \\ \hline
   0.5      & 3   & 0   & 2.7605      & 2.844     \\ \hline
   0.5      & 4   & 0   & 3.6556      & 3.792     \\ \hline
   0.5      & 0   & 1   & 3.9022      & 3.9022     \\ \hline
   0.5      & 1   & 1   & 4.5727      & 4.8682     \\ \hline
   0.5      & 2   & 1   & 5.4033      & 5.7718     \\ \hline
   0.5      & 3   & 1   & 6.2771      & 6.7462     \\ \hline
   0.5      & 0   & 2   & 7.7105      & 7.8044     \\ \hline
   0.5      & 1   & 2   & 8.2279      & 8.7524     \\ \hline
   0.5      & 2   & 2   & 8.9735      & 9.7004     \\ \hline
   0.9      & 1   & 0   & 0.7388      & 0.7388     \\ \hline
   0.9      & 2   & 0   & 1.3294      & 1.4776     \\ \hline
   0.9      & 3   & 0   & 1.8811      & 2.2164     \\ \hline
   0.9      & 4   & 0   & 2.4134      & 2.9552     \\ \hline
   0.9      & 0   & 1   & 3.5994      & 3.5994     \\ \hline
   0.9      & 1   & 1   & 3.9531      & 4.3382     \\ \hline
   0.9      & 2   & 1   & 4.4594      & 5.077     \\ \hline
   0.9      & 3   & 1   & 4.9896      & 5.8158     \\ \hline
   0.9      & 0   & 2   & 6.9420      & 7.988     \\ \hline
   0.9      & 1   & 2   & 7.1733      & 7.9376     \\ \hline
   0.9      & 2   & 2   & 7.6053      & 9.4656     \\ \hline
 \end{tabular}
\caption{Table showing the energy of the vortices for $\nu=2\eta$.}
\label{tab:energies2}
\end{table}

\end{document}